\documentstyle[11pt,epsfig,rotate]{article}
\input{epsfig}
\begin{document}
\def\Tr{\,{\rm Tr}\,}
\def\tr{\,{\rm tr}\,}
\def\beq{\begin{equation}}
\def\eeq{\end{equation}}
\def\beqa{\begin{eqnarray}}
\def\eeqa{\end{eqnarray}}
\begin{titlepage}
\vspace*{-1cm}
\noindent
\phantom{bla}
\\
\vskip 2.0cm
\begin{center}
{\Large {\bf Analysis of ${\cal O}(p^2)$ Corrections to 
$\langle \pi \pi | {\cal Q}_{7,8} | K \rangle$ }}
\end{center}
\vskip 1.5cm
\begin{center}
{\large Vincenzo Cirigliano$^a$ and Eugene Golowich$^b$} \\
\vskip .15cm
$^a$ Dipartimento di Fisica dell'Universit\`a and I.N.F.N. \\
Via Buonarroti,2 56100 Pisa (Italy) \\
vincenzo@het2.physics.umass.edu \\

\vskip .15cm

$^b$ Department of Physics and Astronomy \\
University of Massachusetts \\
Amherst MA 01003 USA\\
gene@het2.physics.umass.edu \\

\vskip .3cm
\end{center}
\vskip 1.5cm
\begin{abstract}
\noindent The one-loop corrections to 
$K \to \pi$ and $K \to 2 \pi$ matrix elements of 
the electroweak penguin operator are calculated.  
General next-to-leading order relations between the 
$K \to \pi$ and $K \to 2 \pi$ amplitudes are obtained.  
The fractional shift $\Delta_2 = 0.27 \pm 0.27$ 
is found for the ${\cal O}(p^2)$ corrections to a recent 
chiral determination of $\langle (\pi\pi)_{I=2}| {\cal Q}_{7,8} 
| K^0 \rangle$. We explain why the sign for $\Delta_2$ is 
opposite to that expected from unitarization approaches based 
on the Omn\`es equation.  We perform a background-field, 
heat-kernel determination of the divergent one-loop amplitudes 
for a general class of (V-A)$\times$(V+A) operators.
\end{abstract}
\vfill
\end{titlepage}

\section{Introduction}
In the Standard Model, $K \to \pi \pi$ amplitudes are conveniently 
expressed in terms of the effective nonleptonic $\Delta S = 1$ 
hamiltonian,
\beq
\langle \pi\pi | {\cal H}_{\Delta S = 1} | K \rangle = 
{G_F \over \sqrt{2}} 
V_{ud} V_{us}^* \sum_{i=1}^{10} \ c_i (\mu) ~
\langle \pi\pi | {\cal Q}_i | K \rangle_\mu \ \ , 
\label{b1}
\eeq
where $\{ {\cal Q}_i \}$ are local four-quark operators, 
$\{ c_i (\mu) \}$ are constants, and $\mu$ is a 
renormalization scale.  In order to determine the 
$K \to \pi\pi$ amplitudes, one must be able to calculate 
the matrix elements $\langle \pi\pi | {\cal Q}_i | K \rangle_\mu $.      
The study of such matrix elements has 
continued to be an active research area because of what it can 
teach us about the inner workings of low energy QCD.  Interest 
has been recently heightened by the KTeV and NA48 
announcements~\cite{ktev} that $\epsilon'/\epsilon \simeq 
20 \cdot 10^{-4}$, to be compared to the approximate theoretical 
relation 
\beqa
& & {\epsilon' \over \epsilon} = 10 \times 10^{-4} \bigg[ 
-3.1~{\rm GeV}^{-3} \cdot \langle (\pi\pi)_{I=0}|{\cal Q}_6
|K^0\rangle 
\nonumber \\
& & \phantom{xxxxxxx} 
- 0.51~{\rm GeV}^{-3} \cdot \langle (\pi\pi)_{I=2}| 
{\cal Q}_8 | K^0 \rangle \bigg] \ \ ,
\eeqa
evaluated in the NDR scheme at scale $\mu = 2$~GeV.  

In the application of chiral symmetry to this problem, the $K \to 
\pi\pi$ matrix elements are expanded in powers of the external
momenta and quark masses.  It is especially attractive to 
consider matrix elements which do not vanish in the 
chiral limit, such as those of the electroweak penguin operators 
\beqa
& & {\cal Q}_7\ = \ {3 \over 2} 
\left( {\bar s}_i~d_i \right)_{\rm V-A} \sum_{q = u,d,s} ~ Q_q
\left( {\bar q}_j~q_j \right)_{\rm V+A} \ \ ,
\nonumber \\
& & {\cal Q}_8\ = \ {3 \over 2} 
\left( {\bar s}_i~d_j \right)_{\rm V-A} \sum_{q = u,d,s} ~ Q_q
\left( {\bar q}_j~q_i \right)_{\rm V+A} \ \ ,
\label{b1a}
\eeqa
where $Q_q$ is the electric charge of quark $q$ and 
$i,j$ are color labels.  Working in the chiral limit, 
Donoghue and Golowich~\cite{dg} recently evaluated the leading chiral 
component to $\langle (\pi\pi)_{I=2}| {\cal Q}_8 | K^0 \rangle$
and also $\langle (\pi\pi)_{I=2}| {\cal Q}_7 | K^0 \rangle$. 
In this paper, we use chiral perturbation theory (ChPT) to 
extend this by calculating the ${\cal O}(p^2)$ chiral corrections to 
matrix elements of ${\cal Q}_7$ and ${\cal Q}_8$.

Given the flavor and chiral structure of ${\cal Q}_{7,8}$, there 
exists a unique operator at ${\cal O} (p^0)$ that represents them 
in an effective chiral description, 
\beq {\cal O}_{\rm ewp} = g \,
 \mbox{Tr} \, \left[ \lambda_{6} U Q U^{\dagger} \right] \ \ .
\label{v1}
\eeq
The operator ${\cal O}_{\rm ewp}$ in turn belongs 
to a family of chiral operators ${\cal O}_{ab}$ which transform 
as members of ${\bf 8}_{L} \times {\bf 8}_{R}$ under chiral rotations, 
\beq 
{\cal O}_{a b} =
 \mbox{Tr} \, \left[ \lambda_{a} U \lambda_{b} U^{\dagger} \right] \ \ .
\label{v2}
\eeq
In Eqs.~(\ref{v1}),(\ref{v2}), $\lambda_{a}$ is a Gell Mann 
matrix, $Q = ~diag~(2/3,-1/3,-1/3)$ is the quark charge matrix 
and $U$ is the matrix of light pseudoscalar fields,  
\beq
U \equiv \exp(i \lambda_k \Phi_k /F) \ \ (k = 1,\dots,8) \ \ ,
\label{c2}
\eeq
where $F$ is the pseudoscalar meson decay constant in lowest order. 
The coupling constant $g$ in Eq.~(\ref{v1}) will depend on the `parent'
operator (${\cal Q}_{7,8}$ in our case) and can be obtained by comparison
with an evaluation of the matrix elements performed in the chiral
limit (analytically as in Ref.~\cite{dg} or QCD lattice-theoretic as in 
Ref.~\cite{Donini:1999nn}).  

The standard first step in a ChPT analysis is to work in the chiral 
world of zero momentum and vanishing light quark mass.  
This leading chiral component, if nonzero, 
often provides the largest contribution to the matrix element.  
In the chiral limit, moreover, simple linear relations exist among 
amplitudes with differing numbers of Goldstone modes.  In particular, 
knowledge of the $K \rightarrow \pi$ matrix elements 
is sufficient to extract the physical $K \rightarrow \pi \pi$ 
matrix elements.  

\begin{figure}
\vskip .1cm
\hskip 1.0cm
\psfig{figure=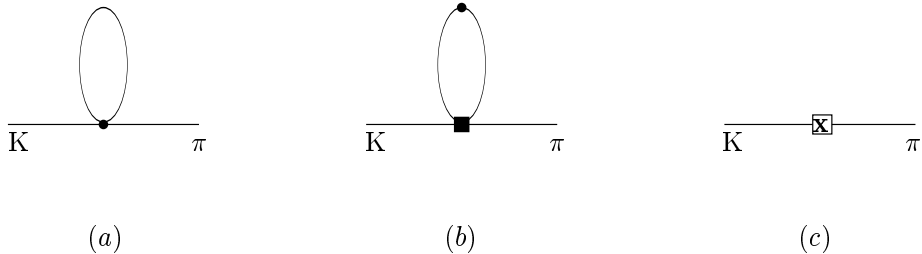,height=1.3in}
\caption{$K \to \pi$ contributions: (a) weak vertex, 
(b) strong and weak vertices, (c) counterterm.\hfill 
\label{fig:loopfig1}}
\end{figure}

Experience has shown, however, the necessity for calculating 
chiral corrections.  This is especially true for amplitudes 
involving kaons, as the possibility of sizeable 
chiral corrections (of order $25 \%  $ or more) cannot be excluded 
when passing from $m_K = 0$ to $m_K = 0.495$~GeV.~\cite{bsw}    
This calls for an ${\cal O}(p^2)$ analysis in the 
chiral representation of weak operators. At this order the chiral 
determination of $K \rightarrow n \pi$ matrix elements of 
${\cal Q}_{7,8}$ includes two basic ingredients:
\begin{enumerate}
\item One-loop diagrams with insertion of vertices from  
${\cal O}_{\rm ewp}$ ({\it cf} Eq.~(\ref{v1})). We calculate 
in dimensional regularization, using a scale parameter 
$\mu_\chi$. 
\item Tree-level diagrams with insertion of the ${\cal O} (p^2)$
counterterm operators (to be described in Sect.~2). 
\end{enumerate}
In the $K \to \pi$ sector, for example, the ${\cal O}(p^2)$ 
amplitudes arise from the contributions depicted in 
Figure~\ref{fig:loopfig1}.\footnote{In 
Figs.~\ref{fig:loopfig1},\ref{fig:loopfig2} strong 
and weak vertices are denoted respectively by large 
bold squares and small bold circles.}
In Fig.~\ref{fig:loopfig1}(a), a loop begins and ends 
at the electroweak vertex, whereas in  
Fig.~\ref{fig:loopfig1}(b) the electroweak vertex is an 
insertion in the loop which is produced by a strong 
interaction vertex.

An important component of our calculation will be a leading-log 
estimate of the ${\cal O} (p^2)$ corrections 
({\it cf}~Eq.~(\ref{result})).  
This provides only a partial estimate of the full 
${\cal O} (p^2)$ effect because it leaves out the 
contribution from the finite counterterms.  Recall 
that divergences encountered in the loop diagrams are 
cancelled by divergences in the counterterm coefficents, 
leaving just the finite loop and counterterm amplitudes.  
Each of these is dependent on the scale $\mu_\chi$, but 
the total amplitude is scale-independent.  Unfortunately, 
the finite counterterm coefficients are unknown at this 
time.  Therefore we base our estimation of ${\cal O} (p^2)$ 
corrections on the finite part of the loop amplitude 
and determine the accompanying uncertainty by 
numerically studying its dependence on $\mu_\chi$. 

In practice, first-principle calculations tend to reliably 
handle only the relatively simple $ K \rightarrow \pi$ matrix 
elements.  It is not clear that this information is 
sufficient to recover matrix elements for multipion final 
states.  Indeed, reconstruction of the physical $ K \rightarrow 
\pi \pi $ matrix elements is known to be 
nontrivial at next-to-leading order.\footnote{See 
Refs.~\cite{bdspw,b,bpp} for the analogous treatment 
of the ${\bf 8}_{L} \times {\bf 1}_{R}$ and 
${\bf {27}}_{L} \times {\bf 1}_{R}$ operators.}
An advantage of the ChPT approach is that it 
provides independent calculations of the $K \to \pi$ and 
$K \to \pi\pi$ transitions. A significant aspect of our investigation 
will be to check whether the the ${\cal O} (p^2)$ 
$K \rightarrow \pi$ matrix elements are sufficient to fully 
determine the the ${\cal O} (p^2)$ $ K \rightarrow \pi \pi $ matrix 
elements.  
 
Our presentation is arranged as follows. 
In Section~2 we list the ${\cal O} (p^2)$ chiral operators representing
${\cal Q}_{7,8}$ and then perform the one-loop renormalizaztion 
of the generating
functional with insertions of the generic operator ${\cal O}_{a b}$ 
(see Eq.~(\ref{v2})). In Section~3 we present the ${\cal O} (p^2)$ 
analysis for $\langle \pi | {\cal Q}_{7,8} | K \rangle $ and 
$\langle \pi \pi | {\cal Q}_{7,8} | K \rangle $, including loop 
and counterterm contributions. Then, in Section~4 we perform a first 
numerical evaluation (based on chiral
loops only) of the corrections to the leading chiral component 
of $\langle \pi \pi | {\cal O}_{7,8} | K \rangle $.  The main
results of this work are summarized and future studies are
outlined in Section~5.  Finally, we note that throughout 
the paper $K \to n \pi$ matrix 
elements of a local operator ${\cal O}$ are expressed as 
\beq
{\cal M}_{K\to\pi} = \langle \pi 
( p_\pi) | {\cal O} | K(p_k) \rangle \quad {\rm and} \quad 
{\cal M}_{K\to\pi_1\pi_2} = \langle \pi 
( p_1) \pi(p_2)| {\cal O} | K(p_k) \rangle \ \ .
\label{amps}
\eeq

\section{Counterterm Lagrangians}

In the following we first display the 
list of ${\cal O}(p^2)$ counterterms which share the 
chiral properties of the ${\cal O}(p^0)$ operator 
${\cal O}_{\rm ewp}$.  Presentation of the various 
$K \to \pi$ and $K \to \pi\pi$ counterterm matrix elements 
is deferred to later in the paper. 

With each counterterm operator will be associated 
a coefficient which is {\it a priori} arbitrary.  Some 
of these counterterm coefficients need to be singular in 
order to cancel divergences which appear in the 
one-loop amplitudes.  In the second part of this section, 
we employ background field and heat kernel methods to 
determine the set of singular ${\cal O}(p^2)$ coefficients.  

\subsection{The Set of Counterterm Lagrangians}\label{subsect:ctlag}
There are seven effective operators at chiral order $p^2$ 
associated with the operator ${\cal O}_{a b}$ of Eq.~(\ref{v2}), 
\beq
{\cal L}^{ab}_{\rm ct} = 
\sum_{k = 1}^7 \ c_k {\cal O}_k^{(ab)} \ \ , 
\label{ct1}
\eeq
with 
\beqa
{\cal O}_1^{(ab)} &=& \Tr \left[ \lambda_a D_\mu U  
\lambda_b D^\mu U^\dagger \right] \ \ ,
\nonumber \\
{\cal O}_2^{(ab)} &=& \Tr \left[ \lambda_a U D_\mu U^\dagger  
\right] \cdot \Tr \left[ \lambda_b U^\dagger D^\mu U  \right] \ \ ,
\nonumber \\
{\cal O}_3^{(ab)} &=& \Tr \left[ \lambda_a U 
\lambda_b D_\mu U^\dagger  D^\mu U  ~U^\dagger + 
\lambda_a U D_\mu U^\dagger  D^\mu U  ~
\lambda_b U^\dagger  \right] \ \ ,
\nonumber \\
{\cal O}_4^{(ab)} &=& \Tr \left[ \lambda_a U \lambda_b S U^\dagger 
+ \lambda_a U S \lambda_b U^\dagger  \right] \ \ ,
\nonumber \\
{\cal O}_5^{(ab)} &=& i \Tr \left[ \lambda_a U [ P, \lambda_b ] 
U^\dagger  \right] \ \ ,
\nonumber \\
{\cal O}_6^{(ab)} &=& \Tr \left[ \lambda_a U \lambda_b U^\dagger 
\right] \cdot \Tr \left[ \chi U^\dagger + \chi^\dagger U \right] \ \ ,
\nonumber \\
{\cal O}_7^{(ab)} &=& \Tr \left[ \lambda_a U \lambda_b U^\dagger 
\right] \cdot \Tr \left[ D_\mu U D^\mu U^\dagger \right] \ \ ,
\label{ct2}
\eeqa  
where the quantities $S$ and $P$ occurring respectively in 
${\cal O}_4^{(ab)}$ and ${\cal O}_5^{(ab)}$ are defined as   
\beq 
S \equiv U^\dagger \chi + \chi^\dagger U 
\qquad  {\rm and} \qquad P \equiv i \left( 
U^\dagger \chi - \chi^\dagger U \right) \ .
\label{ct3}
\eeq
The possible two-derivative dependence $D_\mu D^\mu U$ has been 
eliminated by using the equations of motion, 
\beq
D^\mu \left( U^\dagger D_\mu U \right) + 
{1\over 2} \left( \chi^\dagger U - U^\dagger \chi \right) = 0 \ \ .
\label{ct4}
\eeq
The case of interest for this paper (involving the operator 
${\cal O}_{\rm ewp}$ of Eq.~(\ref{v1})) is recovered with the 
replacements $\lambda_a \to \lambda_6$ and $\lambda_b \to Q$. 
Calculation of the set of counterterm amplitudes is 
straightforward and is summarized in Sect.~\ref{subsect:ct}.

\subsection{Divergences in the One-loop Effective 
Action}\label{subsect:div}
The effective chiral lagrangian on which the one-loop 
analysis is based is 
\beq 
{\cal L} \ = \ {\cal L}^{(2)}_{\rm str} + {\cal O}_{\rm ewp} \ \ ,
\label{k1}
\eeq
where ${\cal O}_{\rm ewp}$ is defined in Eq.~(\ref{v1}) and 
${\cal L}^{(2)}_{\rm str}$ is the familiar 
$\Delta S = 0$ lagrangian,
\beq
{\cal L}^{(2)}_{\rm str} = {F^2_0\over 4}
\Tr \left( D_\mu U D^\mu U^\dagger \right) +
{F^2_0\over 4} \Tr \left(\chi U^\dagger + U\chi ^\dagger\right) \ \ .
\label{k2}
\eeq
We shall calculate quantum corrections about a
solution ${\overline U}$ of the classical theory.  To accomplish 
this, we employ background field and heat kernel 
methods.\footnote{A summary of these techniques appears in 
Appendix~B of Ref.~\cite{dgh}.}  Thus we write
\beq
U \ = \ {\overline U}~e^{i \Delta} \ \ ,
\label{k3}
\eeq
where $\Delta = \lambda^a \Delta^a$ represents the quantum 
fluctuations.  We begin by expressing the generating 
functional at one-loop order as 
\beq
e^{i{\cal Z}} = e^{i \int d^4 x \left[ {\cal L} 
({\overline U}) + {\cal L}_{\rm ct} ({\overline U}) \right] } 
\cdot \int [dU] ~e^{i \int d^4 x ~\left[ {\cal L}(U) - 
{\cal L}({\overline U}) \right] } \ \ ,
\label{k4}
\eeq
where ${\cal L}$ is given in Eq.~(\ref{k1}) and 
${\cal L}_{\rm ct}$ is the counterterm lagrangian 
of Eq.~(\ref{ct1}) with $\lambda_a \to \lambda_6$ 
and $\lambda_b \to Q$. Keeping just the part quadratic 
in $\Delta$, we have 
\beq
{\cal L}(U) - {\cal L}({\overline U}) = - {F^2 \over 2} 
\Delta_a \left( d_\mu d^\mu + \sigma + w \right)^{ab}
\Delta_b + \dots \ \ ,
\label{k5}
\eeq
where 
\beqa
& & d_\mu^{ab} = \delta^{ab} \partial_\mu + \Gamma_\mu^{ab} \ ,
\nonumber \\
& & \Gamma_\mu^{ab} = - {1\over 4} \Tr \left[ 
[\lambda^a , \lambda^b] \left( 
{\overline U}^\dagger \partial_\mu {\overline U} 
+ i {\overline U}^\dagger \ell_\mu {\overline U} 
+ i r_\mu \right) \right] \ ,
\nonumber \\
& & \sigma^{ab} = {1\over 8} \Tr \bigg[ [\lambda^a , \lambda^b]_+ 
\left(\chi U^\dagger + U\chi ^\dagger\right) + 
[\lambda^a , {\overline U}^\dagger \partial_\mu {\overline U}] 
~[\lambda^a , {\overline U}^\dagger \partial^\mu {\overline U}] 
\bigg] \ ,
\nonumber \\
& & w^{ab} = -{g\over F^2} \Tr \bigg[ \lambda_6 {\overline U} 
\left( \lambda^a Q \lambda^b + \lambda^b Q \lambda^a
- {1 \over 2} [[\lambda^a , \lambda^b]_+ , Q]_+ \right) 
{\overline U}^\dagger \bigg]  \ \ . \nonumber \\
\label{k6}
\eeqa
In the above $\ell_\mu$ and $r_\mu$ are source functions 
coupled to chiral currents.  
The integrand occurring in Eq.~(\ref{k4}) is gaussian 
and thus allows for direct evaluation of the path integral, 
\beq
{\cal Z}_{\rm 1-loop} = \int d^4 x \ \left[ {\cal L} 
({\overline U}) + {\cal L}_{\rm ct} ({\overline U}) \right] 
+ {i \over 2} \tr \log \left( d_\mu d^\mu + \sigma + w \right) \ \ ,
\label{k7}
\eeq 
where `$tr$' indicates a sum over spacetime coordinates 
as well as flavor labels.  

By using the heat kernel expansion, we identify the 
divergences in ${\cal Z}_{\rm 1-loop}$ as 
\beq
{\cal Z}_{\rm 1-loop}^{\rm (div)} = {1 \over 2 ( 4 \pi)^{d/2} }
\int d^4 x \ \Gamma\left( {\epsilon \over 2} \right) ~\Tr 
\left[ {1 \over 12} \Gamma_{\mu\nu} \Gamma^{\mu\nu} 
+ {1 \over 2} \left( \sigma + w \right)^2 \right] \ \ ,
\label{k8}
\eeq
where $\epsilon \equiv 4 - d$ and $\Gamma_{\mu\nu}^{ab} = 
[ d_\mu , d_\nu ]^{ab}$.  
There is a great deal of content in Eq.~(\ref{k8}), including 
the one-loop divergent contributions to the Gasser-Leutwyler 
${\cal O}(p^4)$ strong lagrangian.~\cite{gl1,gl2}  
However, we require only the new piece which arises 
from the interference between the quantities $\sigma$ and $w$ 
in Eq.~(\ref{k8}), 
\beq
{\cal Z}_{\rm 1-loop}^{\rm (div)} = {1 \over 2 ( 4 \pi)^{d/2} }
\int d^4 x ~\ \Gamma\left( {\epsilon \over 2} \right) \Tr \left[ 
\sigma ~ w \right] 
= - {{\overline \lambda} \over F^2} 
\int d^4 x ~ \Tr \left[ \sigma ~ w \right] \ \ , 
\label{k9}
\eeq
where ${\overline \lambda}$ is the singular quantity 
\beq
{\overline \lambda} \equiv
{ 1 \over 16 \pi^2} \left[ {1 \over d - 4} - {1\over 2}
\left( \log {4\pi} - \gamma + 1 \right) \right] \ \ .
\label{k9a}
\eeq
Expressing our result in the operator basis (for the case 
$\lambda_a \to \lambda_6$ and $\lambda_b \to Q$)  
of Eq.~(\ref{ct2}), we find for the general case of $N_f$ flavors, 
\beq
{\cal Z}_{\rm 1-loop}^{\rm (div)} = - {{\overline \lambda}g \over F^2} 
\int d^4 x ~ \left[ 2 {\cal O}_2 + {N_f \over 2} \left( {\cal O}_3 
+ {\cal O}_4 \right) + {\cal O}_6 + {\cal O}_7 \right] \ \ .
\label{k10}
\eeq
For three flavors ($N_f = 3$) as in this paper,  
we conclude that the one-loop generating functional will be 
finite provided we use 
\beq
c_i = c_i^{\rm (r)} + {{\overline \lambda}g \over F^2} d_i 
\qquad (i = 1,\dots, 7) \ \ , 
\label{k11}
\eeq
with 
\beq
\begin{array}{l}
d_1 = d_5 = 0 \ , \\
d_2 = 2\ , 
\end{array}
\qquad 
\begin{array}{l}
d_3 = d_4 = 3/2\ ,  \\
d_6 = d_7 = 1 \ \ .
\end{array}
\label{k12}
\eeq

\section{$K \rightarrow \pi$ and $K \rightarrow 2 
\pi$ Amplitudes at ${\cal O}(p^2)$}

When evaluated in the chiral limit, ${\cal O}_{\rm ewp}$ 
has the (tree-level) $K \to \pi$ matrix elements 
\beq
{\cal M}_{K^{+} \to \pi^{+}}^{(0)} = { 2 g \over F^2} 
\ , \qquad 
{\cal M}_{K^{0}\to \pi^{0}}^{(0)} = 0 
\label{finite1.5}
\eeq
and the $K \to 2 \pi$ matrix elements 
\beq
-i {\cal M}_{K^0 \to \pi^+\pi^-}^{(0)} = - 
 {\sqrt{2} g \over F^3}  \ , \qquad  
-i {\cal M}_{K^0\to \pi^0 \pi^0}^{(0)} = 0  \ \ .
\label{finite5.5}
\eeq
These amplitudes have a very simple form, and 
it is not surprising that, as stated in Sect.~1, 
knowledge of the $K \to \pi$ amplitudes is sufficient 
to yield the $K \to 2 \pi$ amplitudes.  We state again 
that the numerical value of the coupling constant $g$ 
can be deduced by referring to the work in 
Refs.~\cite{dg,{Donini:1999nn}}.  However, this information 
is not needed in the present analysis.  

The amplitudes at order $p^2$ will contain both 
loop and counterterm contributions, 
\beq
{\cal M}_i^{(2)} \ = \ {\cal M}_i^{\rm (loop)} \ + \ 
{\cal M}_i ^{\rm (ct)} \ \ ,
\label{tot}
\eeq
which are discussed separately in the subsections to follow.

\subsection{The One-loop Amplitudes}\label{subsect:loop}

Before proceeding to a discussion of the one-loop amplitudes, 
we consider the following technical matter.  In the $K \to \pi$ 
sector, unless the weak operator is allowed to carry off 
nonzero four-momentum, the condition of energy-momentum conservation 
would be valid for physical states only in the SU(3) world 
of degenerate pseudoscalar meson masses 
($p_{\pi}^{2} = p_{K}^{2} \equiv {\bar m}^2$).  
Therefore we shall allow the weak operator to transfer a four-momentum 
$q$. For the $K \to \pi\pi$ matrix elements we can set $q = 0$, but 
we must keep a nonzero value 
of $q$ for the $K \to \pi$ amplitudes.  In particular, 
four-momentum conservation implies $q = p_{K} - p_{\pi}$, where 
$p_{K,\pi}$ are four-momenta of the external kaon and pion.  

We now turn to the $K$-to-$\pi$ matrix elements, for which we provide 
complete analytic expressions.  This allows us to keep track of 
the $q^2$ (or equivalently the $p_{K} \cdot p_{\pi}$) 
dependence,~\cite{rc} and we obtain 
\beqa
& & {\cal M}_{K^{0}\to \pi^{0}}^{\rm (loop)}
= g \bigg[ \frac{1}{\sqrt{2} F^4} \, 
\left[ - {\bar A} (m_{K}^{2})  + I_{K^{0} \pi^{0}} (q^2) \right]  
+ {\cal D}_{K^{0}\to \pi^{0}} \bigg] 
\ , \nonumber \\
& & {\cal M}_{K^{+} \to \pi^{+}}^{\rm (loop)} 
= g \bigg[ \frac{2}{F_{K} F_{\pi}} \, 
    \frac{1}{12 F^2} \left[ \, 34 \, {\bar A} (m_{K}^{2}) + 31 \, 
{\bar A} (m_{\pi}^{2}) + 
      9 \, {\bar A} (m_{\eta}^{2}) \right.   \nonumber \\
& &  \left.  \phantom{xxxxxxxxxxxx} 
+ \, 2 \,  I_{K^{+} \pi^{+}} (q^2)  \right] 
+ {\cal D}_{K^{+}\to \pi^{+}} \bigg] \ \ ,   
\label{finite2} 
\eeqa 
where $g$ is the coupling constant of Eq.~(\ref{v1}).  
In the above expressions, the functions $I_{a b}$ are given by 
\beqa  
& & I_{K^{0} \pi^{0}} (q^2) = (3 p_{K} \cdot p_{\pi} - m_{K}^{2})
{\bar B} (q^2) + 
	{\bar A} (m_{K}^{2}) - \frac{p_{K} \cdot q}{q^2} \, R 
(q^2)  \ ,  \nonumber \\
& & I_{K^{+} \pi^{+}} (q^2) = (2 m_{K}^{2} + m_{\pi}^{2} - 3 p_{K} \cdot 
 	p_{\pi}) \,  {\bar B} (q^2) + {\bar A} (m_{K}^{2}) \nonumber \\ 
& & \phantom{xxxxxxxxxxxxx} 
+ \, \frac{q^2 + 3 (m_{K}^{2} - m_{\pi}^{2})}{q^2} \, R (q^2) \ ,  
\label{finite3}
\eeqa 
where 
\beq  
R (q^2) \equiv (m_{\pi}^{2} - m_{K}^{2}) \, {\bar B} (q^2)  
+ {\bar A} (m_{K}^{2}) - {\bar A} (m_{\pi}^{2}) \ \ , 
\label{finite4}
\eeq 
and ${\bar A} (m^2)$, ${\bar B}(q^2)$ are the {\it regularized} 
one-loop integrals 
\beqa 
& & i {\bar A} (m^2) = \lim_{d=4} \left[ 
2i m^2 {\overline \lambda} + \mu^{4-d}_\chi 
\int {d^{d} l \over (2 \pi)^d} \, \frac{1}{l^2 - m^2}  \right] \ ,
\nonumber \\
& & i {\bar B} (q^2) = \lim_{d=4} \left[ 
2i {\overline \lambda} + \mu^{4-d}_\chi 
\int {d^{d} l \over (2 \pi)^d} \, \frac{1}{(l + q)^2 - m_{K}^{2}} \, 
    \frac{1}{l^2 - m_{\pi}^{2}} \right] \ \ . 
\label{finite5}
\eeqa
The corresponding divergent contributions for 
the $K \rightarrow \pi $ amplitudes, as expressed in 
terms of ${\overline \lambda}$ are 
\beqa
{\cal D}_{K^{0} \to \pi^{0}} & = & \frac{\bar{\lambda}}{\sqrt{2} F^4} \, 
    \left(- 6 \, p_{K} \cdot p_{\pi} \, + \,  2 \, m_{K}^{2} \right)  
    \nonumber \\
{\cal D}_{K^{+}\to \pi^{+}} & = & \frac{\bar{\lambda}}{F^4} \, 
    \left( 2 \, p_{K} \cdot p_{\pi} \, - \, 16 \, m_{K}^{2} \, - \, 
	10 \, m_{\pi}^{2} \right) \ \ .
\label{finite8}
\eeqa

We consider next the two-pion matrix elements. Since one is ultimately 
interested in the physical on-shell result, we set $q = 0$.  This
allows us to replace cumbersome analytic expressions by their 
numerical values.  At one-loop level the physical matrix elements 
will have both real and imaginary parts (as dictated by unitarity). 
Starting with the real parts, we have 
\beq
{\cal R}e \, \left[ -i {\cal M}_i^{\rm (loop)} \right] = g \left[ 
\frac{\sqrt{2}}{F_K F_{\pi}^{2}} \, \left( 
a_i + b_i \, \log \frac{\mu_\chi}{1~{\rm GeV}} \right) 
+ {\cal D}_i \right] \ \ , 
\label{finite7}
\eeq
where the ${\cal D}_i$ are given  
for the $K \rightarrow \pi \pi $ amplitudes by 
\beqa
{\cal D}_{K^0 \to \pi^+ \pi^-} & = & \frac{\bar{\lambda} \sqrt{2}}{F^5} \, 
    \left( \frac{13}{2}  \,  m_{K}^{2}  \,   +  \,  7 \,  m_{\pi}^{2} \right) 
   \ , \nonumber \\
{\cal D}_{K^0 \to \pi^0 \pi^0} & = & \frac{\bar{\lambda} \sqrt{2}}{F^5} \, 
    3 \, \left( m_{K}^{2}-  m_{\pi}^{2} \right) \ \ .
\label{finite9}
\eeqa
The dimensionless coefficients $a_i, b_i$ are collected in Table 1. 

\begin{center}
Table 1 
\end{center}
\begin{center}
\begin{tabular}{ c | c |  c }
\hline\hline
Mode      &  $a_i$   &   $b_i$   \\
\hline  
$~K^0 \to \pi^+ \pi^- $  & $- 1.195$  & $ - 1.300$  \\
$K^0 \to \pi^0 \pi^0 $ & $- 0.654$  &  $- 0.512$  \\
\hline 
\end{tabular}
\end{center}

\noindent Finally, the imaginary parts are found to be 
\beqa
& &  {\cal I}m \, \left[ -i {\cal M}_{K^0 \to \pi^+\pi^-}^{\rm (loop)}
 \right]   = 
- \frac{\sqrt{2} g }{F_K F_{\pi}^{2}} \frac{1}{F^2} 
	\frac{\beta}{16 \pi} \frac{m_{K}^{2}}{2} \ , 
\nonumber \\ 
& & {\cal I}m \, \left[ -i {\cal M}_{K^0 \to 
\pi^0\pi^0}^{\rm (loop)} \right] =  
- \frac{\sqrt{2} g }{F_K F_{\pi}^{2}} \frac{1}{F^2} 
	\frac{\beta}{16 \pi} (m_{K}^{2} - m_{\pi}^{2}) \ ,   
\nonumber \\
\label{finite6}
\eeqa 
where $\beta = (1 - 4 m_\pi^2 /m_K^2 ) ^{1/2}$. 

\subsection{The Counterterm Amplitudes}\label{subsect:ct}

Consider first the $K \to \pi$ transitions, for which we find  
\beqa
{\cal M}_{K^+ \to \pi^+}^{\rm (ct)} &=& {4\over F^2} \bigg[ 
 {1 \over 3} \left(  c_1 - c_3 \right) p_\pi \cdot p_K 
+ \left(  {4 \over 3} c_4 + c_5 + 2 c_6 \right) m_K^2 
\nonumber \\
& & \phantom{xxxxx} + \left( c_4 
+ c_5 + c_6 \right) m_\pi^2 \bigg] \ ,
\nonumber \\
{\cal M}_{K^0 \to \pi^0}^{\rm (ct)} &=& {\sqrt{2} \over F^2} 
\bigg[ \left( {1 \over 3} c_1 + c_2 + { 2\over 3} c_3 \right) 
p_\pi \cdot p_K - {2 \over 3} c_4 m_K^2 \bigg] \ \ ,
\label{ct54}
\eeqa
and next the $K \to \pi \pi$ counterterm amplitudes, 
\beqa
-i {\cal M}_{K^0 \to \pi^+\pi^-}^{\rm (ct)} &=& {\sqrt{2} \over F^3} 
\bigg[  \left(  - {1 \over 3} (c_1 - c_3) - 2 c_4 
- 2 c_5 - 4 c_6 \right) m_K^2 
\nonumber \\
& & \phantom{xxx} + 
\left(  - {2 \over 3} (c_1 - c_3) - 4 c_4 
- 4 c_5 - 2 c_6 \right) m_\pi^2 \bigg] \ ,
\nonumber \\
-i {\cal M}_{K^0 \to \pi^0 \pi^0}^{\rm (ct)} &=& {\sqrt{2} \over F^3} 
\left( m_K^2 - m_\pi^2 \right) 
\bigg[ - {1 \over 3} c_1 - c_2 - { 2\over 3} c_3 \bigg] \ \ .
\label{ct6}
\eeqa
Upon adopting the prescription given in Eqs.~(\ref{k11}),(\ref{k12}), 
the order $p^2$ amplitudes ${\cal M}_i^{(2)}$ are rendered 
ultraviolet finite.  

\subsection{Construction of $K\to \pi\pi$ from $K \to \pi$ 
at ${\cal O}(p^2)$}\label{subsect:constr}

Let us consider now the finite parts of the counterterm amplitudes.  
We shall use the above results 
to show that knowledge of the $K \to \pi$ counterterm amplitudes 
implies knowledge of the $K \to \pi\pi$ 
counterterm amplitudes.  As a result, each $K \to \pi\pi$ 
matrix element becomes expressible in terms of the $K \to \pi\pi$ 
loop amplitude and contributions obtined from the $K \to \pi$ 
sector.  

Recall from Eq.~(\ref{k11}) that we denote the finite parts of the 
counterterm coefficients collectively as $\{ c_i^{(r)} \}$.  
As a first example, we show how to fix the finite counterterm 
$c_4^{(r)}$ which appears as a contribution to 
${\cal M}_{K^0 \to \pi^+\pi^-}^{\rm (ct)}$ in Eq.~(\ref{ct6}).  
From Eq.~(\ref{ct54}), we can in principle obtain $c_4^{(r)}$ by 
taking a partial derivative with respect to the kaon squared-mass of
the counterterm amplitude ${\cal M}^{\rm (ct)}_{K^0 \to \pi^0}$.  
However, one works operationally with the full 
${\cal O}(p^2)$ amplitude 
${\cal M}^{(2)}_{K^0 \to \pi^0}$ and the loop
amplitude  ${\cal M}^{\rm (loop)}_{K^0 \to \pi^0}$.  Thus we have 
\beq
c_4^{(r)} = - {3 F^2 \over 2\sqrt{2}} {\partial \over \partial m_K^2} 
{\cal M}^{\rm (ct)}_{K^0 \to \pi^0} 
= - {3 F^2 \over 2\sqrt{2}} {\partial \over \partial m_K^2} \left( 
{\cal M}^{(2)}_{K^0 \to \pi^0} 
- {\cal M}^{\rm (loop)}_{K^0 \to \pi^0} \right) \ .
\label{ct7}
\eeq
The entire finite counterterm dependence of 
${\cal M}_{K^0 \to \pi^0\pi^0}^{\rm (ct)}$ is 
obtained in like manner, 
\beq
c_1^{(r)} + 3 c_2^{(r)} + 2 c_3^{(r)} = {3 F^2 \over \sqrt{2}} 
{\partial \over \partial (p_\pi \cdot p_K)} \left( 
{\cal M}^{(2)}_{K^0 \to \pi^0} 
- {\cal M}^{\rm (loop)}_{K^0 \to \pi^0} \right) \ \ .
\label{con1}
\eeq
This leaves the set of remaining contributions to 
${\cal M}_{K^0 \to \pi^+\pi^-}^{\rm (ct)}$, for which we find
\beqa
& & c_1^{(r)} - c_3^{(r)} = {3 F^2 \over 4} 
{\partial \over \partial (p_\pi \cdot p_K)} \left( 
{\cal M}^{(2)}_{K^+ \to \pi^+} 
- {\cal M}^{\rm (loop)}_{K^+ \to \pi^+} \right) \ ,
\nonumber \\
& & c_5^{(r)} = {F^2 \over 4} \left[ 
2 {\partial \over \partial m_\pi^2} - 
{\partial \over \partial m_K^2} \right] 
\left( {\cal M}^{(2)}_{K^+ \to \pi^+} 
- {\cal M}^{\rm (loop)}_{K^+ \to \pi^+} \right) 
\nonumber \\
& & \phantom{xxxxxxxxx} + 
{F^2 \over \sqrt{2}} {\partial \over \partial m_K^2} \left( 
{\cal M}^{(2)}_{K^0 \to \pi^0} - {\cal M}^{\rm (loop)}_{K^0 \to 
\pi^0} \right) \ ,
\nonumber \\
& & c_6^{(r)} = {F^2 \over 4} \left[ 
{\partial \over \partial m_K^2} - 
{\partial \over \partial m_\pi^2} \right] 
\left( {\cal M}^{(2)}_{K^+ \to \pi^+} 
- {\cal M}^{\rm (loop)}_{K^+ \to \pi^+} \right) 
\nonumber \\
& & \phantom{xxxxxxxxx} + 
{F^2 \over 2\sqrt{2}} {\partial \over \partial m_K^2} \left( 
{\cal M}^{(2)}_{K^0 \to \pi^0} 
- {\cal M}^{\rm (loop)}_{K^0 \to \pi^0} \right) \ .
\label{con2}
\eeqa
The collection of finite counterterms in 
Eqs.~(\ref{ct7})-(\ref{con2}) is seen to recover 
the entire content of the $K \to \pi\pi$ counterterm amplitudes.  
The appearance of partial derivatives in these equations 
suggests plotting the dependence of the $K\to \pi$ 
amplitudes for each of the kinematic variables $m_K^2$, $m_\pi^2$, 
$p_\pi \cdot p_K$ and numerically extracting the slopes.

\section{A First Estimate of the ${\cal O}(p^2)$ Chiral Corrections}

At this point we have sufficient information to 
perform a first estimate of corrections
to the physical matrix elements away from the chiral limit.
Our estimate will involve only the chiral logarithms and, as 
discussed earlier, is therefore by no means complete. 
In particular, the answer in this approximation 
depends upon the dimensional regularization scale $\mu_{\chi}$.  
An additional limitation is given by the fact that the chiral-log 
approximation does not distinguish between ${\cal Q}_7$ and 
${\cal Q}_8$.  In both cases it gives the same fractional 
correction to the results obtained in the chiral limit. 
We turn now to the calculation of the fractional corrections 
and then compare our findings with those expected from an 
Omn\`es type unitarization approach.

\subsection{The Fractional Shifts $\Delta_0$ and 
$\Delta_2$}\label{subsect:shift}
In order to recover the quantities commonly used in the literature, 
we must take matrix elements between $K$ and a final two-pion state 
of definite isospin.  Denoting $ {\cal M}_{I} = \langle (\pi \pi)_I  | 
{\cal Q}_{7,8} | K \rangle $, we can write each isospin 
amplitude, including chiral loops, as 
$$ {\cal M}_{I} = {\cal M}_{I}^{(0)} \cdot 
\left( 1 + \Delta_I \right) \ ,  $$
where ${\cal M}_{I}^{(0)}$ is evaluated in the 
chiral world.  We determine the central value of the 
${\cal O}(p^2)$ correction $\Delta_I$ by averaging the finite one-loop 
amplitude over the range $\mu_{\chi}: 0.5 \rightarrow 1$ GeV.  The 
error bars in this determination are estimated from the 
range of values observed while varying the chiral scale $\mu_{\chi}$ 
as indicated above.  This uncertainty reflects 
our present ignorance of the low energy constants. 
Our procedure results in the ${\cal O}(p^2)$ corrections, 
\beq
 \Delta_0 =  0.98  \pm  0.55  \ ,  \qquad \qquad 
 \Delta_2  =  0.27 \pm 0.27 \ \ .
\label{result}
\eeq
These values omit possible contributions, of 
unknown magnitude, from the finite counterterms.  
It is clear the counterterm contribution 
cannot be very much smaller than that of the 
loop, else its scale dependence would not undergo 
the necessary cancelation against the loop component.  
In principle it could be much larger, in which case 
corrections exceeding the above ranges would occur.  
Even from our leading-log estimate, however, we see 
that sizeable deviations from the chiral limit predictions 
cannot be excluded and that a more detailed calculation of 
such corrections is called for.

\begin{figure}
\vskip .1cm
\hskip 1.0cm
\psfig{figure=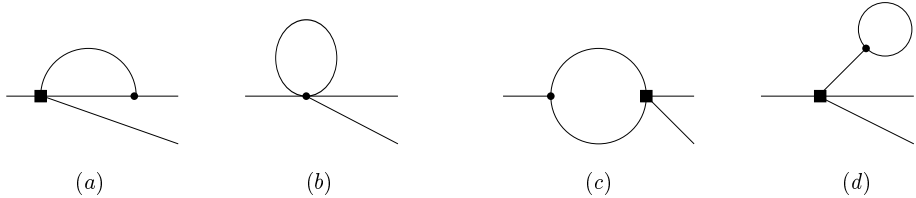,height=1.0in}
\caption{$K \to \pi\pi$ one-loop contributions. \hfill 
\label{fig:loopfig2}}
\end{figure}

\subsection{Comparison with Unitarization 
Procedures}\label{subsect:omnes}
The possible impact of final-state interactions (FSI) on weak 
transition amplitudes has recently drawn attention in the 
literature.~\cite{{pp},{ep}}  Using the work of Ref.~\cite{pp} 
as an example, the $K \to (\pi\pi)_I$ amplitude induced by 
the usual nonleptonic weak interaction is written in the form 
\beq
{\cal A}_I (s) = {\cal A}_I^{(0)} ( s - m_\pi^2) e^{i 
\delta_0^I (s)} {\cal R}_I(s) \ \ , 
\label{fsi1}
\eeq 
where ${\cal A}_I^{(0)}$ is the derivative of ${\cal A}_I$ 
at the point $s = m_\pi^2$ and ${\cal R}_I(s)$ is a dispersive 
corection factor.  The above form is an Omn\`es-type 
solution which includes the effect of final-state rescattering.
In the approximation of including just the two-pion final states, 
Ref.~\cite{pp} obtains for the $I = 0,2$ rescattering corrections  
\beq
{\cal R}_0 (m_K^2) = 1.41 \pm 0.06 \ , \qquad 
{\cal R}_2 (m_K^2) = 0.92 \pm 0.02 \ \ .
\label{fsi2}
\eeq  
This means that, within errors, the effect of final state 
interactions enhances the $I=0$ amplitude by $41\%$ and 
suppresses the $I=2$ amplitude by $8\%$.  
It is revealing to compare these FSI effects with our 
${\cal O}(p^2)$ fractional corrections 
$\Delta_0$ and $\Delta_2$ given in 
Eq.~(\ref{result}).  Our finding that $\Delta_0$ 
is large and positive is in qualitative accord 
with the above value for ${\cal R}_0 (m_K^2)$, but 
our positive value for $\Delta_2$ conflicts 
with the suppression implied by ${\cal R}_2 (m_K^2)$.

The resolution of this apparent paradox lies in a more 
careful comparison of the two approaches.  In 
Fig.~\ref{fig:loopfig2}(a)-(d), we display the four 
distinct one-loop contributions to our $K \to \pi\pi$ 
electroweak penguin matrix elements.  The contribution of 
Fig.~\ref{fig:loopfig2}(c) (the `back-facing swordfish') 
is the only one with a two-pion intermediate state and 
is the only contribution whose iteration would occur in 
an Omn\`es-type resummation.  A calculation of the 
fractional shift induced by Fig.~\ref{fig:loopfig2}(c) 
yields 
\beq
 \Delta_0^{2(c)} =  0.36  \pm  0.12  \ ,  \qquad \qquad 
 \Delta_2^{2(c)}  =  - 0.15 \pm 0.05 \ \ .
\label{fsi3}
\eeq
Both these values are in reasonable agreement with those implied 
by Eq.~(\ref{fsi2}).  It is the {\it other} contributions in 
Fig.~\ref{fig:loopfig2} (which we emphasize are demanded 
by the stringent requirements of chiral symmetry) which are 
dominant and which determine the overall sign of $\Delta_2$.  

\section{Conclusion}

The calculation of matrix elements of local operators which enter the 
$ \Delta S = 1$ weak effective hamiltonian is 
a crucial step in the phenomenology of nonleptonic kaon decays. 
In this paper we have focussed on some general aspects of the 
${\cal O} (p^2)$ corrections to the 
electroweak penguin operators ${\cal Q}_{7,8}$. 
Our results can be summarized as follows:
\begin{enumerate} 
\item Direct calculational methods (such as lattice-QCD, 
chiral sum rules,
large-$N_{c}$ QCD) usually deal with the simpler 
$K \rightarrow \pi$ transitions and rely on chiral symmetry 
relations to infer the physical $K \rightarrow \pi \pi$ matrix
elements.  Although trivial in leading order (which is 
${\cal O} (p^0)$ in this case), these relations become more involved 
at next-to-leading order in the chiral expansion.  We have
systematically worked out the form of the $K \rightarrow \pi$ and $K
\rightarrow \pi \pi$ matrix elements at ${\cal O} (p^2)$ in the
framework of ChPT, including one-loop diagrams and local counterterms 
(as required by power counting). We find that
knowledge of the $K \rightarrow \pi$ matrix elements {\it is} 
generally enough to fully determine the low energy constants entering 
in the $K \rightarrow \pi \pi$ matrix elements. Thus {\it all} 
chiral studies of this system, analytic or lattice-QCD, can 
safely use the $K \to \pi$ mode to infer properties of the 
$K \to \pi\pi$ system.
\item We have used background-field and heat-kernel techniques to 
perform the renormalization of the one-loop effective action 
in the presence of an insertion of the generic 
${\bf 8}_L \times {\bf 8}_R$ operators ${\cal O}_{a b}$ of
Eq.~(\ref{v2}).  Although the results given in 
Eqs.~(\ref{k10}),(\ref{k12}) pertain specifically to the 
electroweak penguin operator 
of Eq.~(\ref{v1}), the procedure is easily generalized to 
applications involving other left-right operators.  
The divergent counterterm coefficients predicted by 
the general heat-kernel approach are in complete agreement 
with the divergences explicitly found in our one-loop 
calculation and thus provide a check on the calculation.

\item Another important aspect of our work is the `chiral-loop' 
estimate of the ${\cal O} (p^2)$ correction to $\langle 
(\pi \pi)_I | {\cal Q}_{7,8} | K \rangle 
$ ({\it cf}~Eq.~(\ref{result})).  On a qualitative 
level, we have answered the question 
of whether the corrections might be anomalously small 
(${\cal O}(m_\pi^2)$) due to the absence of ${\cal O}(m_K^2)$ 
contributions.  For example, recall that electromagnetic corrections 
to the $K^+ \rightarrow \pi^+ \pi^0$ matrix element 
of the octet weak lagrangian contains only small 
${\cal O}(\alpha m_{\pi}^2)$ corrections even though the 
larger ${\cal O}(\alpha m_{K}^2)$ terms are allowed by power 
counting and symmetry arguments.  We have found here that 
${\cal O}(m_K^2)$ effects are not absent.  This is why the 
leading-log result exhibits a strong dependence on the chiral 
scale $\mu_{\chi}$.  Our analysis cannot, therefore, exclude 
the possibility of important corrections to the chiral determination.  

More quantitatively, we obtain the large ${\cal O}(p^2)$ 
correction $\Delta_0 = 0.98 \pm 0.55$ in the isospin-zero 
two-pion channel.  It is not surprising that this agrees with 
Omn\`es-type resummation procedures~\cite{{pp},{ep}} 
because strong rescattering effects are expected to play 
an important role for the $I=0$ channel.  Of greater 
interest is the $I=2$ two-pion channel, since it 
bears on attempts to predict $\epsilon' / \epsilon$.  
Here, the situation is more subtle.  The fractional correction 
$ \Delta_2 = 0.27 \pm 0.27$ is moderate in magnitude and 
positive in sign.  The sign of $\Delta_2$ is opposite to 
that obtained in an Omn\`es-type resummation approach. As explained 
in Sect.~\ref{subsect:omnes} this is attributable to ${\cal O}(p^2)$ 
contributions not included in the unitarized amplitudes.  The 
underlying lesson is that unitarization of a part of the 
full amplitude can lead to valid results when rescattering is 
strong but can be misleading when rescattering is weak.  
\end{enumerate}

This work represents a preliminary but necessary step towards 
a fully predictive analysis of the matrix elements $\langle \pi \pi 
| {\cal Q}_{7,8} | K \rangle$ beyond the chiral limit. 
Work is underway to extend the dispersive analysis carried out 
in Ref.~\cite{dg} to the general case of $m_{\rm quark} \neq 0$.

\vspace{1.cm}

The research described here was supported in part by the 
National Science Foundation.  One of us (V.C.) acknowledges 
support from M.U.R.S.T.  We thank John Donoghue for useful  
comments on this work.

\eject

\end{document}